\documentstyle{article}

\begin{document}
\begin{flushleft}
{\it Foundations of Physics Letters, Vol.\ 11, No.\ 4, pp.\ 359--369, 1998}
\end{flushleft}              

\section*{COMMON MISREPRESENTATION 
          OF THE EINSTEIN--PODOLSKY--ROSEN ARGUMENT\\}
\noindent
\hspace*{10ex}{\bf V. Hnizdo}\footnote[0]{Address since July 1999: National
Institute for Occupational Safety and Health, 1095 Willowdale Road,
Morgantown, WV 26505, USA; e-mail: vhnizdo@cdc.gov}\\
\newline
\hspace*{10ex}{\it Department of Physics,
                   Schonland Research Centre for Nuclear}\\
\hspace*{10ex}{\it Sciences, and Centre for Nonlinear Studies}\\
\hspace*{10ex}{\it University of the Witwatersrand, Johannesburg,
                   2050 South Africa}\\
\newline
\hspace*{10ex}{\rm Received 23 January 1998; revised 27 June 1998}\\
\newline
A frequently given version of the argument of Einstein, Podolsky and Rosen
against the completeness of the quantum mechanical description
is criticized as a misrepresentation that lacks the cogency of the original
EPR argument.\\
\newline
Key words: Einstein--Podolsky--Rosen argument, quantum measurements,
physical reality, completeness of quantum mechanics.
\newline
\section*{1. INTRODUCTION}
\noindent The argument of Einstein, Podolsky and Rosen (EPR),
advanced by these authors in support of the thesis that quantum mechanics
cannot provide a complete description of a physical system,
has been receiving close attention in foundational studies and debates
on quantum mechanics ever since the famous paper of EPR appeared
in 1935 [1].
Not lacking in subtlety, the EPR argument was bound to be
misrepresented frequently in the extensive writings on the subject.
A particular version of the EPR argument, analyzed critically as a
misrepresentation already more than 25 years ago by Hooker [2],
who classified it as ``the third related EPR argument," has
been appearing persistently in the literature, especially in that aimed
at a more general audience,\footnote[1]{For example, even in the
admirable
analysis and interpretation of the philosophy of N. Bohr by Folse [3].}
and in popular accounts.\footnote[2]{See, for example, Ref.\ [4].}
The present paper is occasioned by a clearly stated occurrence
of this version in a book on the relatively recent but already
influential ``consistent-histories," or ``logical," revision [5--7]
of the Copenhagen interpretation of quantum mechanics, written by one of its
originators R.~Omn{\`e}s [8].

The important difference between the original EPR argument and,
to use Hooker's classification, ``the third related" version
will be pointed out using the presentation of Omn{\`e}s;
moreover, it will be argued anew that ``the third related" version
is a misrepresentation that loses the cogency of the argument as it was
put forward originally.\footnote[3]{Of course, the EPR argument is compelling
only when their concept of reality is accepted, see  Secs.\ 2 and 3.}
This will agree with the findings of Hooker [2], but for different
reasons than those he advanced; Hooker's reasons will be criticized
as flawed on the physical grounds.
\newline
\section*{2. THE ORIGINAL EPR ARGUMENT}
\noindent
As Omn{\`e}s says himself in his book, nothing can replace the direct
reading of the original paper of Einstein, Podolsky and Rosen
when studying their argument against the completeness of the
quantum mechanical description. Following this advice should not be an
onerous task as this famous paper has only four pages, in which
the argument is developed clearly and succinctly---and the reader is
entreated to peruse the paper to gain a first-hand knowledge
of the matter under the present discussion. The following three paragraphs
are a synopsis of the EPR argument, with comments that attempt
to facilitate an appreciation of the intuitive physical context
behind the formal logical structure of the argument.

A satisfactory theory should be complete, and a necessary condition for
the completeness of a theory is that it has a counterpart for every
``element of physical reality." A sufficient criterion for recognizing
an element of physical reality that corresponds to some physical quantity
is satisfied when one can predict with certainty the value of that
quantity without in any way disturbing the system to which the quantity
pertains. Such a criterion aims at defining rigorously an
empirically discernable symptom of the existence of physical
reality that is independent of the observer and his measurements.
For a {\it conserved} physical quantity,
an intuitive justification for this criterion can be given easily
along the following lines:
if one can, {\it without in any way disturbing} the given system, predict
the value of a conserved quantity pertaining to the system, the quantity
must have had that value from the time of the last external disturbance of
the system, and it will keep that value as long as the system
remains undisturbed, in particular before any measurement is performed on
the system, including a measurement of the quantity concerned.
Generally, considering also nonconserving quantities, the EPR reality
criterion is a condition that one would accept intuitively as
guaranteeing, when it is fulfilled, that a measurement of the quantity
concerned only {\it reveals} the quantity's value, which exists
independently of whether the measurement is performed or not.

Having formulated their demands on a complete theory in terms of
elements of physical reality that can be identified as such according to a
well-defined criterion, EPR note that the wavefunction $\Psi(x_1,x_2)$
of a system consisting of two noninteracting subsystems that, however,
interacted with each other in the past, in general can be expanded in terms
of complete sets $u_n(x_1)$ and $v_s(x_1)$ of eigenfunctions
corresponding to different physical quantities $A$ and $B$, respectively,
of system 1:
\begin{equation}
\Psi(x_1,x_2)=\sum_n \psi_n(x_2)u_n(x_1)=\sum_s\phi_s(x_2)v_s(x_1).
\end{equation}
Here, the functions $\psi_n(x_2)$ and $\phi_s(x_2)$ can be regarded as the
coefficients in the two expansions.
The summations above run over at least two terms, and so the state (1)
is now often called ``entangled,"\footnote[4]{This terminology goes back
to Schr{\"o}dinger [9].}
in the sense that it does not factorize into a single product of two
functions where each depends only on the degrees of freedom of one of the
subsystems. EPR point out that one can assign different wavefunctions
to subsystem 2 by performing different
local measurements on subsystem 1. This is because of what they term
``the reduction of the wave packet" upon a measurement, now usually called
the collapse of the wavefunction: when the quantity $A$ is measured on
subsystem 1 and found to have a value $a_k$, subsystem 1 is left in a
state described by the wavefunction $u_k(x_1)$, collapsing the
wavefunction (1) into a single term $\psi_k(x_2)u_k(x_1)$,
which means that subsystem 2 is left in a state described
(to within a constant) by a wavefunction $\psi_k(x_2)$; but if the
quantity $B$, instead of $A$, had been chosen
to be measured and a value $b_r$ obtained, the state of the system would
have collapsed to the state $\phi_r(x_2)v_r(x_1)$, leaving subsystem 2
in a different state $\phi_r(x_2)$.

The two wavefunctions $\psi_k$ and $\phi_r$ may happen to be eigenfunctions
of two noncommuting operators corresponding to conjugate physical quantities
$P$ and $Q$, respectively. EPR illustrate this by considering a two-particle
system prepared in a state in which, along a given direction,
the difference of the position coordinates of the particles and the sum of
their momenta have sharp values
$x_0$ and $0$, respectively:\footnote[5]{As observed immediately by Bohr
in his reply to EPR [10], such a state is perfectly legitimate in quantum
mechanics because the operators of the difference in
the positions and of the sum of the momenta of two particles commute.}
\begin{equation}
\Psi(x_1,x_2)=\delta(x_2-x_1-x_0).
\end{equation}
For simplicity, they assume only a one-dimensional motion, and use a
non-normalizable delta function instead of a more
realistic wave packet that would have a finite spread in position and
momentum.
The above entangled state can be written as
\begin{equation}
\Psi(x_1,x_2)=\int \psi_p(x_2)u_p(x_1)\,dp,
\end{equation}
where
\begin{equation}
u_p(x_1)=\frac{1}{2\pi\hbar}\,\exp({\rm i}px_1/\hbar)\;\;{\rm and}\;\;
\psi_p(x_2)=\exp[-{\rm i}p(x_2-x_0)/\hbar]
\end{equation}
are momentum eigenfunctions of particles 1 and 2 with eigenvalues
$p$ and $-p$, respectively; it can be written also as
\begin{equation}
\Psi(x_1,x_2)=\int \phi_x(x_2)v_x(x_1)\,dx,
\end{equation}
where
\begin{equation}
v_x(x_1)=\delta(x_1-x)\;\;{\rm and}\;\;\phi_x(x_2)=\delta(x_2-x-x_0),
\end{equation}
are position eigenfunctions of particles 1 and 2 with eigenvalues
$x$ and $x+x_0$, respectively.
Now when a
measurement of the momentum of particle 1 yields a value $p$,
one knows with certainty that the momentum of particle 2 must be
$P=-p$, because on that measurement the wavefunction (3) collapses to
the state $\psi_p(x_2)u_p(x_1)$, where $\psi_p(x_2)$, the state
in which particle 2 is left, is the eigenfunction in (4) of $P$ with
the eigenvalue $-p$.
But one is free to measure the position of particle 1, instead of its
momentum. If it was then
decided to measure the former, on the resulting value $x$ of such a
measurement one could have predicted with certainty
that the position of particle 2 is $Q=x_0+x$, because the wavefunction (5)
would have collapsed to the state $\phi_x(x_2)v_x(x_1)$, where $\phi_x(x_2)$,
the state of particle 2, is the eigenfunction in (6)
of $Q$ with the eigenvalue $x_0+x$.
EPR assert that the momentum $P$ and position $Q$ of particle 2
must possess simultaneous reality according to their reality
criterion, as the predictions of $P$ and $Q$ can be done by measurements
on the noninteracting particle 1 and thus without disturbing particle 2
in any way.\footnote[6]{This can be ensured by having the particle separation
$x_0$ sufficiently great; curiously, EPR do not mention explicitly this
simple measure.}
The formalism of quantum mechanics, however, does not allow a
state in which a particle has simultaneously a definite momentum and a
definite position, and so EPR conclude that the quantum-mechanical
description is not complete.
While EPR admit that one cannot predict the position and momentum
of particle 2 simultaneously, they counter this objection by asserting
that the reality of the position and momentum of particle 2 cannot, in any
``reasonable definition of reality," depend on the process of measurement
on particle 1 carried out without disturbing particle 2.
In other words, they argue that the simultaneous
reality of the position and momentum of particle~2 is established already
by the {\it possibility} of predicting either of the two quantities on the
result of a measurement of {\it either} the position {\it or} momentum
of the noninteracting particle~1. In drawing this conclusion, they
use not only their reality criterion, but also rely on their firm conviction
that in any ``reasonable" concept of physical reality,
the real state of a system must be independent of what may happen
to another system, from which the system in question is well separated
and with which it has no interaction.

In 1951, Bohm [11] employed in the EPR argument
a more convenient system of two spin-1/2 particles prepared in the entangled
state of zero total spin $J$,
\begin{equation}
\textstyle
|JM\rangle=|00\rangle=\sqrt{\frac{1}{2}}\bigg[
 {\left|\frac{1}{2}\right\rangle}^{(1)}_{\bf\hat n}
 {\left|-\frac{1}{2}\right\rangle}^{(2)}_{\bf\hat n}
-{\left|-\frac{1}{2}\right\rangle}^{(1)}_{\bf\hat n}
{\left|\frac{1}{2}\right\rangle}^{(2)}_{\bf\hat n}\bigg],
\end{equation}
where ${\left|\pm\frac{1}{2}\right\rangle}^{(i)}_{\bf\hat n}$ is the spin
state of particle $i$ with the projection of its spin on the axis along an
arbitrary unit vector $\bf\hat n$ equal to $\pm\frac{1}{2}$. The particles
are assumed to be far away from each other; for example, they could
be the $s$-wave products of the decay of a quasibound two-spin-1/2-particle
system of zero total angular momentum.
Since its introduction, the Bohm's version has been used almost exclusively
in the literature as it has the advantage of using
quantities that do not have continuous spectra and are conserved,
thus freeing the discussion from non-essential aspects such as those
connected with the non-normalizability of wavefunctions and the
time evolution of the system, which may distract from the gist of the
argument. Using the state (7), the EPR argument runs then as follows.
The $z$-component $s_{2z}$ of the spin of particle 2 is an element of reality
because it can be predicted, without disturbing particle 2 in any way, by
measuring the $z$-component $s_{1z}$ of the spin of the distant particle 1:
if such a measurement yields $s_{1z}=+\frac{1}{2}$ (or $-\frac{1}{2}$),
it follows from the zero-total-spin state (7), with $\bf\hat n=\hat z$,
that $s_{2z}$ must then equal $-\frac{1}{2}$ (or $+\frac{1}{2}$).
But the $x$-component $s_{2x}$ of the spin of particle
2 must be an element of reality, too, as one could predict it similarly by
performing an alternative measurement of the $x$-component $s_{1x}$ of
the spin of particle 1 and using $\bf\hat n=\hat x$ in (7).
The quantum-mechanical description then must be incomplete as it does
not allow a state in which both the $z$-component and $x$-component of the
spin of a particle have definite values.

A comment on the role of the collapse of the wavefunction
in the EPR argument seems now in order. While EPR talk about the
collapse of the wavefunctions (1), (3) and (5) into single products of
wavefunctions of subsystems 1 and 2 as a result of suitable measurements
on subsystem 1, it can be seen easily that
the collapse of the wavefunction does not need to be evoked
explicitly in order to carry the argument through.
The {\it joint} observables of the difference in the positions
and of the sum of the momenta of two particles
can be each measured by measuring the relevant observables on each of the
two particles separately: the particles' individual positions for the
first joint observable, and their individual momenta for the second
one.\footnote[7]{However, such measurements of joint observables cannot
be used to prepare the composite system in an entangled state like
the one of Eq.\ (2); an experimental procedure that would prepare such
an entangled state has been described by Bohr [10]. Moreover, such
measurements, even when they are of the repeatable type, would disentangle
a given entangled state of the composite system---and this is made use of
in the argument of the rest of this section.}
But as the state (2) is an eigenstate of both the joint observables,
the result of the measurement of the position or the momentum
of particle 1, together with the knowledge of the eigenvalues of
the joint observables in the state (2), i.e.,
the position difference $x_0$ and the momentum sum 0,
enables one to predict the value of the position or momentum of particle 2
already on the universally accepted meaning of an eigenstate and its
eigenvalues---namely that, in such a state, a measurement of the
observable (which is here of the joint type) pertaining to an eigenvalue of
the state must yield with certainty that eigenvalue.
Similarly, no collapse of the zero-total-spin state (7) is needed explicitly
in the argument, only the fact that the total spin is zero and thus
$s_{1z}+s_{2z}=s_{1x}+s_{2x}=0$.
Of course, it may be useful to couch the discussion in terms
of suitable wavefunction collapses, but such collapses can be deduced here
from the above mentioned meaning of an eigenstate and its eigenvalues with
no need of postulating them as logically primary.\footnote[8]{
Here it is noted that the ``no-collapse measurements"
proposed by Home and Whitaker [12], which are supposed to leave
a system always in the state that the system had
before measurement, can be ruled out using a similar argument.
Indeed, if an entangled state of a composite system remained unchanged
after the measurement of a suitable observable on one of its two subsystems,
a subsequent measurement of the corresponding observable on the other
subsystem would not necessarily yield a value that
results in the eigenvalue of the relevant joint observable in
the entangled state when the value is combined with the measurement result
on the first subsystem.}
\newline
\section*{3. ``THE THIRD RELATED" ARGUMENT}
\noindent
The EPR argument employs measurements, performed or contemplated, on
{\it only one of the two subsystems}, say particle 1, of the system.
So far as particle 2 is concerned, its state is not directly measured,
it is rather predicted on the result of a direct measurement on particle 1.
However, the presentation of the EPR argument by Omn{\`e}s employs
measurements on {\it both} particles: in the Bohm's version,
a measurement of the $z$-component $s_{1z}$ of particle 1 {\it and}
a subsequent measurement (say at a time $t$) of the spin of particle 2
along another direction, say a measurement of the $x$-component $s_{2x}$.
On the result of the measurement of $s_{1z}$ one can predict
the value of the $z$-component $s_{2z}$ of the spin of particle 2 and thus,
according to the EPR reality criterion, the value of $s_{2z}$ of particle 2
is an element of physical reality, which, obviously, it will remain to be so
until the time $t$ of the direct measurement of $s_{2x}$.
But the latter measurement assigns to particle 2 a definite value
of $s_{2x}$, and so, at the time $t$, particle 2 must have definite values
of both $s_{2z}$ and $s_{2x}$, which is in direct conflict
with the possibilities of the quantum-mechanical description.
There is no fundamental reason, apart from that of convenience,
for not performing the above measurements of $s_{1z}$ and $s_{2x}$
simultaneously; in fact, when the original example with position
and momentum quantities is used in this version of the EPR argument,
the position of one of the particles and the momentum of the other
should be measured simultaneously
to circumvent complications due to the time evolution of the system.
Versions of the EPR argument that employ  measurements
on both the two subsystems belong to the class identified by
Hooker and labeled by him as ``the third related EPR argument."\footnote[9]
{A recent coincidence-measurement presentation of the EPR argument by
Domingos {\it et al.} [13] thus falls in this class also;
this paper has been commented on critically in Ref.\ [14].}

The essential aspect in which ``the third related" and  original
EPR arguments differ is that the former aims at establishing the
possibility of simultaneous {\it determination} of two conjugate
quantities, whereas in the latter it is argued for simultaneous
{\it reality} (or {\it existence}) of such quantities.
As is well known, one can find easily measurement procedures that assign
simultaneously arbitrarily precise values to position and momentum, but they
all have a retrospective character in the sense that they never prepare
a state in which these values are the initial conditions.\footnote[10]
{Heisenberg discussed such retrospective measurements on an example of the
time-of-flight method of velocity measurement already in 1930 [15],
and Bohr commented on such a method of velocity measurement even earlier,
in his historic Como paper (Ref.\ [16], p.\ 66 in {\it Atomic Theory and the
Description of Nature}). A more recent discussion
of retrospective measurements can be found in Ref.\ [17].}
The uncertainty principle refers to the preparation of a state, which has
testable consequences, rather than to a retrospective measurement
with no predictive power. Einstein was well aware of that,
and his failed pre-EPR attempts at disproving the uncertainty principle
were aimed at finding experimental procedures that would lead to
a simultaneous determination of two conjugate quantities in such a way
that it would have a predictive power.\footnote[11]{Einstein's famous
clock-in-the-box thought experiment is perhaps the best example, see
Ref.\ [18].}
The EPR argument differs from the previous attempts of Einstein in not
trying to show that one can determine simultaneously two conjugate
quantities; EPR must have realized that simultaneous measurements of
the position of one of the particles and the momentum of the other cannot
lead to a determination with predictive power of the simultaneous position
and momentum of any of the two particles. In fact, EPR steer carefully from
what would be ``the third related argument": while they accept that the two
conjugate quantities of particle 2 cannot be predicted at the same
time because only one of the two alternative measurements on the correlated
particle 1 can be performed actually, they do not counter this by proposing
to perform the other measurement on particle 2  (this would be
``the third related argument").
EPR argue ``only" for simultaneous reality of the two conjugate quantities,
using, with no need to perform in the end any measurements,
their reality criterion in conjunction with their firm belief
that the {\it real} state of a system cannot depend
upon the process of local measurement performed on another system,
well separated from the first system so that the measurement cannot disturb
the latter.
Bohr clearly understood that well,
and aimed his reply to EPR accordingly at their concept of reality [10].

``The third related" EPR argument is thus not only a misrepresentation
of the original EPR argument, but also lacks cogency:
the simultaneous determination of two conjugate quantities it
purports to establish has no predictive power; it cannot lead to
a physical state in which the values of two conjugate quantities are
among its initial conditions and the knowledge of which would enable one
to predict the future time development of the given system.
To use the original EPR example with the conjugate quantities
of position and momentum, a momentum measurement on particle 2
results in a loss of its spatial coordination with respect
to the frame of reference, as demanded by the uncertainty principle,
and the correlation it had with the position of particle 1, which could be
used to determine its position, is thus lost irretrievably.
Similarly, concerning now particle 1, its position measurement
destroys irrecoverably the correlation it had with the momentum
of particle 2
and which could be used to determine its momentum on the result
of the momentum measurement on particle 2.\footnote[12]
{Bohr stresses this point in his reply to EPR [10].}
The prediction of the value of a quantity pertaining to
a given particle on the result of a measurement of the similar quantity
on the other particle,
together with the simultaneous measurement of the conjugate quantity
on the given particle,
thus gives the simultaneous values of the two conjugate quantities
of the given particle only in the retrospective sense.

Hooker in his excellent study of the Einstein--Bohr debate argues
also against the cogency of ``the third related" EPR argument,\footnote[13]
{See Ref.\ [2], Sec.\ 5 and pp. 223--224.} but for reasons that
are criticized presently as flawed. Hooker asserts that in fact one
cannot perform sufficiently accurately the simultaneous measurements
of the position of particle 1 and the momentum of particle 2
because there is, as Bohr has shown [10,18],
an uncontrollable transfer of momentum
to the common coordinate frame as a result of the position measurement 
on particle 1. While it is certainly true that an accurate position
measurement must lead to an uncontrollable transfer of momentum to the
body that serves as a coordinate frame of reference, this circumstance
presents no practical difficulty when the frame is sufficiently massive.
Indeed, the momentum that a sufficiently massive frame absorbs in the
position measurement of particle 1 results in the frame acquiring
a negligible velocity, and thus the space-time coordination
of any auxiliary body that is used in the determination of the momentum
of particle 2 is not affected.\footnote[14]{A comment that has no direct
bearing
on the issues at hand seems nevertheless worth making here. To achieve
a position determination of such an auxiliary body without affecting its
momentum, Hooker proposes to use a parallel beam of light to illuminate a
position pointer attached to the body, with a light detector placed behind
the pointer, see Ref.\ [2], pp.\ 215--217. It can be seen easily, however,
that also in such a procedure there will be an uncontrollable transfer of
momentum to the body on account of the diffraction of light by the pointer's
edge. Such diffraction determines the ultimate accuracy of the position
measurement and, as it involves an uncontrollable change in the direction
of the incident photons, it leads to an uncontrollable transfer of momentum
to the body along the direction in which its position is measured.
Fortunately, the momentum uncertainty introduced by a position
measurement does not affect the accuracy of the velocity determination
by the time-of-flight method, in the context of which Hooker
makes the above proposal, if the time interval
between the two position measurements needed in such a method
is sufficiently great [15].}
As Bohr always stressed [18], proper space-time coordination uses
a frame and clocks that can absorb the inevitable and uncontrollable
transfers of momentum and energy without being affected in their role as
a space-time framework. Thus, contrary to Hooker's assertion, the position
of particle 1 and the momentum of particle 2 can be measured simultaneously
and, in principle, with arbitrary accuracy.
However, the assignment of simultaneous values to conjugate quantities
of any of the two particles, to which such measurements lead in a procedure
that follows ``the third related" EPR argument,
amounts simply to a retrospective measurement with no testable consequences.
\newline
\section*{4. CONCLUDING SUMMARY}
\noindent
A frequently given version of the EPR argument against the completeness
of the quantum mechanical description, called by Hooker ``the third
related" EPR argument, was criticized as a misrepresentation
of the original argument. The cogency of ``the third related" argument
itself was disputed on the grounds that it aims at establishing
the possibility of a determination of simultaneous values
of conjugate quantities, rather, as in the original EPR argument, at
establishing their simultaneous reality, and that as such it amounts only
to a scheme of a retrospective measurement. 
\newline
\section*{REFERENCES}
\begin{enumerate}
\item A. Einstein, B. Podolsky, and N. Rosen,
``Can quantum-mechanical description of physical reality be considered
complete?,"
{\it Phys.\ Rev.} {\bf 47}, 777--780 (1935); reprinted in J. A. Wheeler and
W. H. Zurek, eds., {\it Quantum Theory and Measurement} (Princeton University
Press, Princeton, New Jersey, 1983), pp.\ 138--141.
\item C. A. Hooker, ``The nature of quantum mechanical reality:
Einstein versus Bohr," in R. G. Colodny, ed., {\it Paradigms and Paradoxes}
(University of Pittsburgh Press, Pittsburgh, 1972), pp.\ 67--302.
\item H. J. Folse, {\it The Philosophy of Niels Bohr:
The Framework of Complementarity} (North-Holland, Amsterdam, 1985),
pp.\ 146--154.
\item  M. White and J. Gribbin, {\it Einstein: A Life in Science}
(Simon and Schuster, London, 1994), Chap.\ 12, p.\ 218.
\item R. B. Griffiths, ``Consistent histories and the interpretation
of quantum mechanics," {\it J.\ Stat.\ Phys.} {\bf 36}, 219--272 (1984).
\item M. Gell-Mann and J. B. Hartle, in W. H. Zurek, ed., {\it Complexity,
Entropy, and the Physics of Information} (Santa Fe Institute
Studies in the Science of Complexity, No.\ 8) (Addison-Wesley, Redwood City,
California, 1991).
\item R. Omn{\`e}s,``Consistent interpretations of quantum mechanics,"
{\it Rev.\ Mod.\ Phys.} {\bf 64}, 339--382 (1992); ``A new interpretation
of quantum mechanics and its consequences in epistemology," 
{\it Found.\ Phys.} {\bf 25}, 605--629 (1995).
\item R. Omn{\`e}s, {\it The Interpretation of Quantum Mechanics}
(Princeton University Press, Princeton, New Jersey, 1994), Chap.\ 9.
\item E. Schr{\"o}dinger,
``Die gegenw{\"a}rtige Situation in der Quantenmechanik," {\it
Naturwiss.} {\bf 23}, 807--812; 823--828; 844--849 (1935); English
translation
as ``The present situation in quantum mechanics," in J. A. Wheeler and
W. H. Zurek, {\it op.\ cit.}, pp. 152--168.
\item N. Bohr, ``Can quantum-mechanical description of physical 
reality be considered complete?,"
{\it Phys.\ Rev.} {\bf 48}, 696--702 (1935); reprinted in
J. A. Wheeler and W. H. Zurek, {\it op.\ cit.}, pp.\ 145--151.
\item D. Bohm, {\it Quantum Theory} (Prentice-Hall, Engelwood Cliffs,
New Jersey, 1951), pp. 611--623; reprinted in J. A. Wheeler and W. H. Zurek,
{\it op.\ cit.}, pp.\ 356--368.
\item D. Home and M. A. B. Whitaker, ``Interpretations of quantum 
measurement without the collapse postulate," {\it Phys.\ Lett.\ A} {\bf 128}, 
1--4 (1988).
\item J. M. Domingos, F. Nogueira, M. H. Caldeira and F. D. dos Aidos,
``EPR: Copenhagen interpretation has got what it takes,"
{\it Eur.\ J.\ Phys.} {\bf 17}, 125--130 (1996).
\item V. Hnizdo, ``EPR and the Copenhagen interpretation,"
{\it Eur.\ J.\ Phys.} {\bf 18}, 404--406 (1997).
\item W. Heisenberg, {\it The Physical Principles of the Quantum 
Theory} (Dover, New York, 1930), p.\ 25.
\item N. Bohr, ``The quantum postulate and the recent development
of atomic theory," {\it Nature} {\bf 121}, 580--590 (1928); reprinted
in N. Bohr, {\it Atomic Theory and the Description of Nature}
(Cambridge University Press, Cambridge, 1934 and 1961), pp.\ 52--91;
reprinted also in J. A. Wheeler and W. H. Zurek, {\it op.\ cit.},
pp.\ 87--126.
\item L. E. Ballentine, ``The statistical interpretation of quantum
mechanics," {\it Rev.\ Mod.\ Phys.} {\bf 42}, 358--381 (1972).
\item N. Bohr, ``Discussions with Einstein on epistemological problems
in atomic physics," in P. A. Schlipp, ed., {\it Albert Einstein:
Philosopher-Scientist} (Open Court, Evanston,
Illinois, 1949), pp.\ 199--241; reprinted in N. Bohr, {\it Atomic Physics and
Human Knowledge} (Wiley, New York, 1958), p.\ 32; reprinted also in
J. A. Wheeler and W. H. Zurek, {\it op.\ cit.}, pp.\ 9--49.

\end{enumerate}

\end{document}